# Remarks on the Relevance of Privacy Expectations for Default Opt-out Settings

Sebastian Zimmeck
*Department of Mathematics and*
*Computer Science*
*Wesleyan University*
Middletown, CT, United States
szimmeck@wesleyan.edu

*Abstract*—Over the past few years an increasing number of states in the US have adopted new privacy laws. The majority of these laws require compliance with universal opt-out mechanisms (UOOMs), which allow consumers to send legally binding opt-out signals. However, a number of laws generally do not allow UOOMs to be enabled by default. While some laws exempt privacy-protective software from this prohibition, the exemption does not apply to pre-installed software, e.g., a privacy-protective web browser bundled with an operating system. The reason for not allowing default opt-out settings for pre-installed software is to ensure that settings reflect consumers' "affirmative, freely given, and unambiguous choice," as, for example, the Colorado Privacy Act (CPA) is putting it. However, prohibiting vendors of privacy-protective software from turning on UOOMs by default can force them into committing unfair or deceptive acts or practices under the FTC Act and equivalent state laws. Thus, whether UOOMs can be turned on by default on pre-installed software should depend on consumers' privacy expectations. For pre-installed software that is creating a reasonable expectation for consumers that their privacy will be protected, the simple use of such software should be considered a valid choice for enabling UOOMs. In such software a turned-on UOOM is not a "default setting" but rather the software's inherent behavior that a consumer expects and chooses through its use. This interpretation of consumer choice is preferable under the CPA and similar laws as it grounds the notice and choice principle in the privacy expectations of consumers and enables companies to compete on better privacy for consumers.

I. INTRODUCTION

The privacy landscape in the US is changing. Over the past few years an increasing number of states have adopted new privacy laws (Figure 1) [1]. The majority of these laws require compliance with universal opt-out mechanisms (UOOMs), which allow consumers to send legally binding signals to opt out of the sale or sharing of their personal information or to opt out of targeted advertising [2]. Global Privacy Control (GPC) is one example of such a UOOM [3]. However, various laws, for example, the Colorado Privacy Act (CPA), generally do not allow browser and other software vendors to enable UOOMs by default [4]. While there is an exemption for privacy-protective software, it only applies if it is user-installed [5]. It does not apply to pre-installed software, such as a privacy-protective web browser that comes bundled with an operating system. As an example, consider Rule 5.04 of the CPA Rules:

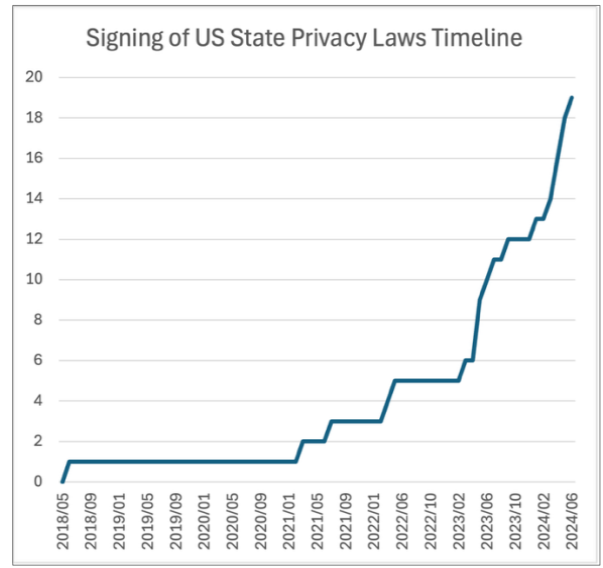

Fig. 1. The signing of US state privacy laws from June 2018, when the California Consumer Privacy Act (CCPA) was signed into law, until June 2024. Activity picked up substantially in early 2021.

*[A] Universal Opt-Out Mechanism may not be the default setting for a tool that comes pre-installed with a device, such as a browser or operating system.*

*Example: An operating system manufacturer bundles a browser pre-installed with every device shipped with the operating system. The browser sends a Universal Opt-Out mechanism signal by default and never asks the Consumer to enable this setting. The Consumer's decision to use this browser does not represent the Consumer's affirmative, freely given, and unambiguous choice to use the Universal Opt-Out Mechanism because it is a default choice. This is so even if the marketing for the operating system touts its privacy-protective features.*

The result of Rule 5.04 of the CPA Rules and similar laws is that some browser vendors are allowed to turn on UOOMs by default while others are not. Even more so, the same browser is treated differently depending on whether it comes pre-installed.

The rationale for not allowing pre-installed software to turn on UOOMs by default is to ensure that consumers' settings reflect their "affirmative, freely given, and unambiguous choice," CPA, Section 6-1-1313(2)(c). Turning on a UOOM by default could be the choice of the software vendor and not the choice of the consumer, especially, since most people do not change their privacy default settings [6]. Since the US is an opt-out regime, if there is no indication that consumers want to opt out, the default will be that they stay opted in. Thus, consumers need to engage in some activity that indicates that they want to opt out. Doing so generally requires privacy labor [7]. Thus, consumers would need to weigh exercising their opt-out rights against the amount of work they have to perform. This consideration may decrease the use of UOOMs and the very reason for their existence as they are, in fact, intended to make it easier for consumers to exercise their opt-out rights. The more difficult it is for consumers to opt out, the more will not exercise their right.

## II. A Brief History of UOOM Defaults

Not allowing opt-out settings being enabled by default has its origins in the Do Not Track (DNT) effort that began nearly twenty years ago. In 2007 various consumer and privacy organizations urged the Federal Trade Commission (FTC) to protect consumers from online behavioral tracking and targeting by creating a national Do Not Track List similar to the national Do Not Call List [8]. Based on this idea privacy advocates Chris Soghoian and Sid Stamm implemented a simple Firefox plug-in to add a DNT header to the requests a browser sends to a server [9]. In 2010 the FTC took up this idea and recommended a browser-based mechanism through which consumers can express a single, persistent preference of whether they allow targeted advertising [10]. Then, Microsoft, whose Internet Explorer had major market share at the time, added DNT to its browser [11]. Meanwhile, companies, consumer organizations, researchers, and other web platform stakeholders had begun work in the World Wide Web Consortium (W3C) Tracking Protection Working Group to establish a standard for DNT [12]. In its 2012 final report, "Protecting Consumer Privacy in an Era of Rapid Change," the FTC noted that industry has made significant progress in implementing DNT, however, added that more work remains to be done [13]. At that point, Microsoft made a crucial decision that would prove fateful for DNT.

For its most recent version of Internet Explorer, Internet Explorer 10, Microsoft announced that it would turn on DNT by default [14]. Users would not need to change a setting to tell websites that they do not want to be tracked, but rather the browser would come preset with DNT. Immediately after this decision Microsoft faced criticism from many companies in the ad industry criticizing that turning on DNT should be an explicit user choice and not a browser vendor choice [15]. Critics also pointed out that Microsoft's decision would violate the Digital Advertising Alliance's (DAAs) agreement with the US government, which only required honoring DNT signals if those were not sent by default [15]. Accordingly, the DAA advised companies that they are "not require[d] […] to honor DNT signals fixed by the browser manufacturers and set by them in browsers." They further explained that "it is not a DAA Principle or in any way a requirement under the DAA Program to honor a DNT signal that is automatically set in IE10 or any other browser" [16]. Different from Microsoft, Mozilla did not turn on DNT by default in Firefox as the signal should represent "a choice made by the person behind the keyboard and not the software maker, because ultimately it's not Firefox being tracked, it's the user" [17]. Based on this reasoning, Roy Fielding, co-editor of the DNT standard, wrote a patch for the Apache web server to disable DNT for requests coming from Internet Explorer 10 and entitled it "Apache does not tolerate deliberate abuse of open standards" [18]. While the patch was reversed a few weeks later [19], the damage was already done.

At this point, it is worthwhile to take a step back and consider how DNT actually works. DNT signals—just as other UOOM signals—are not enforced at the software level. They merely express a preference upon which the recipient must act by implementing technologies that prevent personal information sharing, selling, and targeting. Thus, the enforcement of UOOMs relies on regulatory agencies' investigation of companies' compliance. The web does not technically enforce UOOM compliance. To that end, DNT relied (and still does) on California Assembly Bill (AB) 370 [20], which was signed into law in 2013 and became effective on January 1, 2014. AB 370 amended the California Online Privacy Protection Act (CalOPPA) by requiring an operator of a commercial web site or online service to "[d]isclose how the operator responds to Web browser 'do not track' signals or other mechanisms that provide consumers the ability to exercise choice regarding the collection of personally identifiable information about an individual consumer's online activities over time and across third-party Web sites or online services," CalOPPA, 22575(b)(5). Notably, CalOPPA does not require operators of online services to respect DNT; it only requires them to disclose whether they do. Thus, most operators' standard practice is to simply disclose, in friendly terms, in their privacy policies that they do not respect DNT. That is all that CalOPPA requires. Lawmakers' hopes with AB 370 that the online ad industry would regulate itself proved to be illusory.

Despite the failure of DNT, its underlying idea, a simple user agent setting for opting out, is compelling. Thus, it has been revived in the majority of US state privacy laws that lawmakers enacted in recent years: California, Colorado, Connecticut, and other states' privacy laws contain opt-out rights that consumers can exercise via UOOMs. DNT serves as a valuable lesson for this second attempt of automating consumer choice. Most importantly, we have learned that self-regulation does not work in this context and that it will only marginally change the status quo, if at all.

## III. Evolving Notice and Choice For "Defaults"

The online advertising industry is built upon a complex and deeply integrated system that is inherently resistant to change. Individual participants are not able to make major changes on their own and are constrained by the system in which they operate. Thus, mandatory UOOM compliance serves as a vital regulatory lever for improving consumer privacy in the online advertising industry. In this context we should take what we learned from the DNT default debate and avoid drawing the wrong conclusion, i.e., that a software vendor should be categorically prohibited from turning on a UOOM by default on pre-installed software.

The question of defaults is a question of the notice and choice principle, according to which consumers' make an active choice based on the privacy notice they receive. However, a consumer's choice is not limited to toggling a software setting. A choice can also be implicit. As Aleecia McDonald already pointed out during the DNT deliberations [21]: "the act of selecting [a] user agent is itself a choice that expresses the user's preference for privacy." More broadly, a consumer's choice can be manifested through the selection of a product or service that is advertised to be privacy-protective of its users. That is the notion underlying the CCPA: "[t]he consumer exercises their choice by affirmatively choosing the privacy control, including when utilizing privacy-by-design products or services" [22].

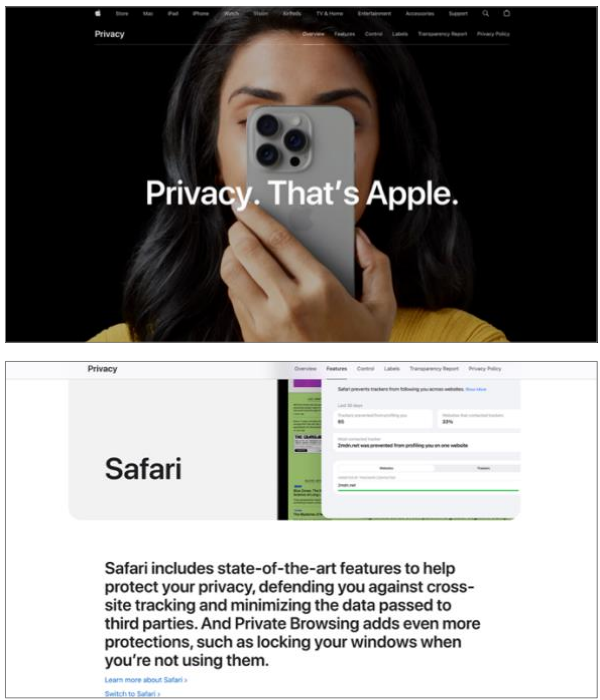

Fig. 2. Screenshots of Apple's privacy page (top) [25] and Apple's privacy features page (bottom) [26] as of October 26, 2025.

The notice to consumers of a product's or service's privacy-protective nature can be conveyed in marketing materials that the company provides. One notable example is Apple, which is constructing its brand identity and market differentiation strategy on protecting the privacy of its users. It achieves this positioning in the market through marketing campaigns, public statements, and deliberate privacy design. Consequently, it is through these acts that Apple cultivates an expectation in the mind of consumers that their privacy is protected when they use Apple products or services. Users can trust that Apple's products and services—including the pre-installed Safari browser [23]—are privacy-protective.

Thus, if consumers choose an Apple product over a competitor's product that is less steeped into privacy claims, they make a choice that is inextricably linked to Apple's privacy promises creating the expectation that their privacy will be protected. Importantly, such consumer expectations are not just marketing; they have legal relevance as well. Section 5 of the FTC Act (15 USC §45) and its state-level equivalents, Unfair or Deceptive Acts or Practices (UDAP) laws, establish a binding legal obligation on companies to not engage in unfair or deceptive acts or practices.

The test for determining whether an act or practice is deceptive is whether consumers' expectations or interpretations are reasonable in light of the claims made [24]. A representation, omission, or practice is deceptive if it is likely to mislead a consumer acting reasonably under the circumstances and is likely to affect a consumer's conduct or decision regarding a product or service [24]. Thus, when Apple broadly advertises "Privacy. That's Apple." (Figure 2), such advertisement creates a reasonable expectation that its software will protect the privacy of its users by default—even when it comes pre-installed. A browser that is marketed as "includ[ing] state-of-the-art features to help protect […] privacy, defending […] against cross-site tracking and minimizing the data passed to third parties" is at odds with a default setting that allows the sale and sharing of personal information or targeting of its users.

As to unfairness, an act or a practice is unfair if it causes or is likely to cause substantial injury to consumers, cannot be reasonably avoided by consumers, and is not outweighed by countervailing benefits to consumers or to competition [24]. In this regard, the FTC views a violation of a reasonable privacy expectation as a "substantial injury" [27]. It found "data brokers' collection, aggregation, and disclosure of location data [to] violate consumers' expectations of privacy" [27]. In the case of Apple, this injury is not reasonably avoidable because the consumer already took the avoidance step by purchasing the privacy-protective product or service.

Fundamentally, the lens through which the FTC interprets a company's acts and practices to determine whether they are unfair or deceptive are general consumer expectations. The FTC's tests, as well as the tests under state UDAP laws, apply without regard to whether a particular consumer cared about privacy or is using the product or service because of its privacy-protections. The question for whether a consumer was deceived is not "What did a particular consumer expect?" but "What did an ordinary consumer expect?" [28] Similarly, it is not necessary for proving unfair acts or practices to show that a consumer actually relied on a privacy promise [29]. Thus, a company that markets its products and services as privacy-protective yet disables default opt-out settings thereby allowing data selling, sharing, or targeting may commit unfair or deceptive acts or practices. In such case the company should be allowed to turn on opt-out settings by default. Such default opt-outs would be the materialization of consumers' reasonable privacy expectations.

As it stands, the current interpretation of the CPA prohibiting default opt-out settings for pre-installed software conflicts with the FTC Act and state UDAP laws. Given their prohibition of unfair and deceptive acts and practices, the latter *require* companies like Apple to enable UOOMs by default to meet consumer expectations. However, the prevalent CPA interpretation *prohibits* Apple from doing so. In this conflict of laws, a company is forced to comply with one law (the CPA) but doing so would violate others (the FTC Act and state UDAP laws). Under the current interpretation of the CPA, Apple would be forced to prompt consumers to turn on a privacy setting that they can reasonably expect to be already active by virtue of their choice of an Apple service or product.

When state law conflicts with federal law, the latter prevails making the state law void or unenforceable per the Supremacy Clause, Article VI, Clause 2 of the Constitution. In such conflicts federal law preempts state law requiring states to yield to federal statutes, treaties, and the Constitution. In line with this requirement the CPA provides in Section 6-1-1304(3)(a)(I) that the obligations under the CPA "do not restrict a controller's or processor's ability to comply with federal, state, or local laws, rules or regulations." To avoid such a conflict we can extend the interpretation of "choice" from toggling a setting to also cover implicit choices.

CPA, Section 6-1-1313(2)(c) requires software to "not adopt a mechanism that is a default setting, but rather clearly represent[] the consumer's affirmative, freely given, and unambiguous choice to opt out." Fundamentally, the law is based on the dichotomy between choices and default settings, which are mutually exclusive. A default setting does not represent a choice and vice versa. However, for Safari and other privacy-protective software a UOOM being turned on is not a default setting but its inherent nature. Thus, Safari users do not choose to turn on the UOOM but rather their choice manifests in using the browser that inherently functions this way as is. The use of such a privacy-protective product implies a choice. They can still turn off the UOOM. However, the expected behavior of Safari is to protect its users privacy.

This interpretation is consistent with the ban of default opt-out settings as it is intended to prevent scenarios like Microsoft's Internet Explorer enabling DNT by default. A non-privacy-protective browser would impose a UOOM on the consumer if is turned on by default as it is not in the nature of such browser to function this way. The interpretation also aligns with Rule 5.04 of the CPA. While the Rule prohibits a default-on UOOM for pre-installed browsers "even if the marketing for the operating system touts its privacy-protective feature," it does not prohibit such setting if the marketing of the *browser* makes clear that it protects users' privacy.

Certainly, if Apple explains in its marketing materials for Safari that UOOMs are turned on, consumers would have reasonable privacy expectations in this regard. Such explanation would also be in line with California's new AB 566 [30], which requires browsers to have a preference signal setting. AB 566 mandates that, by 2027, a "business that develops or maintains a browser shall make clear to a consumer in its public disclosures how the opt-out preference signal works and the intended effect." If Apple explains in its marketing materials, i.e., its "public disclosures," that UOOMs are turned on by default, it can simultaneously comply with AB 566 and create alignment of Safari's functionality with its marketing under the FTC Act and state UDAP laws.

For the CCPA, the interpretation of choice in light of the privacy-protective nature of a software is explicit. According to the CCPA a "consumer exercises their choice by affirmatively choosing the privacy control, including when utilizing privacy-by-design products or services" [22]. In this regard the CCPA provides a blueprint for the CPA, whose Section 6-1-1313(2)(e) mandates that its opt-out mechanisms be as consistent as possible with similar mechanisms required by other US laws. To achieve this goal, the CPA's interpretation of choice' should be harmonized with the CCPA recognizing that selecting privacy-protective software is, in itself, the exercise of a consumer's choice.

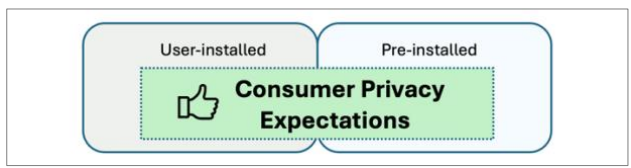

Fig. 3. Consumers can have reasonable privacy expectations in user-installed software but also in pre-installed software. Whether software is user- or pre-installed has no relevance for consumers' privacy expectations.

The central question for determining whether a UOOM can be turned on by default is what a consumer's reasonable privacy expectations are. The focus on the install form as the determining factor for choice is inadequate. The default setting should not depend on whether software is user- or pre-installed but rather on the privacy expectations the vendor creates in its marketing and other disclosures (Figure 3). Thus, a software vendor should be allowed to turn on a UOOM by default to avoid liability under the FTC Act and state UDAP laws due to an unfulfilled reasonable expectation in the minds of consumers that their personal information will not be sold or shared or that they will not be targeted with ads.

## IV. ENABLING COMPANIES TO COMPETE ON PRIVACY

Beyond conflicts in the application of laws, a rule of not allowing privacy-protective default settings for pre-installed privacy-protective software also hinders competition for privacy. The prohibition of default opt-out settings on pre-installed software not only frustrates consumer expectations but also makes pre-installed software appear less privacy-protective than user-installed software, which can enable UOOMs by default. Such a strict reading disadvantages privacy-by-design software simply because it is pre-installed thereby making it more burdensome for consumers to use pre-installed privacy-protective software. More generally, automating privacy features, notably by defaults, plays an important role to reduce consumer privacy transaction costs and support the development of a market for privacy [31].

## V. CONCLUSIONS

UOOMs appear popular with consumers [32, 33]. Whether they can be turned on by default is a question of balancing explicitness with usability. Explicitly toggling an opt-out setting has a low degree of ambiguity but makes a product less usable. On the other hand, an implicit opt-out via a use decision may be more ambiguous but has higher usability. The test for where to draw this line should not be based on the question of whether software was pre-installed but the central privacy expectation question: what reasonable privacy expectations do consumers have?

When viewed through this lens, the principle becomes clearer. For example, for Apple enabling a UOOM by default in privacy-protective software is not a vendor-imposed setting. Rather, the use of the browser, including this setting, is a choice consumers can be expected to make for a browser that inherently behaves privacy-protective. This interpretation of the CCPA avoids unfair and deceptive practices under the FTC Act and state UDAPs. Conversely, for a vendor whose business model relies on data collection and who makes no privacy promises, a default-on UOOM would be an overreach that the law should (and does) prohibit.

Ultimately, the default question has implications beyond UOOMs as it touches the core of the notice and choice principle. Daniel Solove has argued that the principle should be replaced by substantive, non-waivable privacy rights [34]. Whether or not that should be the case requires discussion, but what we can say here is that a broader understanding of notice and choice beyond toggling settings offers a more realistic and usable experience of privacy choice for consumers. Especially, when consumers pay a premium for "Privacy. That's Apple." they should get privacy, by default.

ACKNOWLEDGMENT

I thank my reviewer for the improvement suggestions. I am grateful to the National Science Foundation for their support of this research (Award #2055196). I also would like to thank Wesleyan University, its Department of Mathematics and Computer Science, and the Anil Fernando Endowment for their additional support. Conclusions reached or positions taken are my own and not necessarily those of my supporters, its trustees, officers, or staff.